\documentstyle[12pt]{JHEP3}

\newcommand{\e}{\epsilon}
\renewcommand{\t}{\theta}
\newcommand{\tb}{\overline{\theta}}
\renewcommand{\a}{\alpha}
\newcommand{\s}{\psi}
\renewcommand{\c}{\chi}

\newcommand{\be}[1]{ \begin{equation}\label{#1} }
\newcommand{\ee}{\end{equation}}
\newcommand{\ben}[1]{\begin{eqnarray}\label{#1} }
\newcommand{\een}{\end{eqnarray}}
\newcommand{\eq}[1]{(\ref{#1})}
\def\ZZZ{{\hskip-3pt\hbox{ Z\kern-1.6mm Z}}}
\def\zzz{{\hskip-3pt\hbox{ z\kern-1mm z}}}

\newcommand{\p}{\partial}

\newcommand{\Q}{{\mathcal{Q}}}
\renewcommand{\S}{{\mathcal{S}}}
\newcommand{\A}{{\mathcal{A}}}

\newcommand{\refb}[1]{(\ref{#1})}

\def\one{{\hbox{ 1\kern-.8mm l}}}
\def\zero{{\hbox{ 0\kern-1.5mm 0}}}

\preprint{HRI/ST/0913}

\title{Supersymmetric Extension of Galilean Conformal Algebras}

\author{
Arjun Bagchi, Ipsita Mandal \\
$\;$ Harish-Chandra Research Institute,\\
$\;$ Chhatnag Road, Jhusi, Allahabad 211019, INDIA\\
$\;$\email{arjun, ipsita@hri.res.in}
}

\abstract{The Galilean conformal algebra has recently been realised in the study of the non-relativistic limit of the AdS/CFT conjecture. This was obtained by a systematic parametric group contraction of the parent relativistic conformal field theory. In this paper, we extend the analysis to include supersymmetry. We work at the level of the co-ordinates in superspace to construct the $N=1$ Super Galilean conformal algebra. One of the interesting outcomes of the analysis is that one is able to naturally extend the finite algebra to an infinite one. This looks structurally similar to the $N=1$ superconformal algebra in two dimensions, but is different. We also comment on the extension of our construction to cases of higher $N$.}

\begin{document}

\baselineskip 3.5ex

\section{Introduction}

The non-relativistic versions of the AdS/CFT conjecture \cite{Maldacena:1997re} have recently received a lot of attention. The motivation has mainly been studying real-life systems in condensed matter physics via the gauge-gravity duality\footnote{See \cite{Hartnoll:2009sz} for a recent review of AdS/Condensed Matter Theory (AdS/CMT) correspondence.}. It was pointed out in \cite{Nishida:2007pj} that the Schr\"{o}dinger symmetry group \cite{Hagen:1972pd, Niederer:1972zz,Henkel:1993sg}, a non-relativistic version of conformal symmetry, is relevant to the study of cold atoms. A gravity dual possessing these symmetries was then proposed in \cite{Son:2008ye, Balasubramanian:2008dm}. 

Recently, the study of the actual non-relativistic limit of the conjecture was initiated in \cite{Bagchi:2009my}, where the authors proposed to study a non-relativistic conformal symmetry obtained by a parametric contraction of the relativistic conformal group. One of the additional motivations for this is that there might be possibly interesting tractable sectors of the parent conjecture, like the BMN limit \cite{BMN}, which emerge when we look at such a non-relativistic limit. The process of group contraction of the relativistic conformal group $SO(d+1,2)$ in $d+1$ space-time dimensions\cite{Lukierski:2005xy}, leads in $d=3$ to a fifteen parameter group (like the parent $SO(4,2)$ group) which contains the ten parameter Galilean subgroup. This Galilean conformal group is to be contrasted with the twelve parameter Schr\"{o}dinger group (plus central extension) with which it has in common only the non-centrally extended Galilean subgroup. The Galilean conformal group is different from the Schr\"{o}dinger group in some crucial respects. For instance, the dilatation generator $\tilde{D}$  in the Schrodinger group scales space and time differently 
$x_i \rightarrow \lambda x_i, t\rightarrow \lambda^2 t$. Whereas the corresponding generator $D$ in the Galilean Conformal Algebra (GCA) scales space and time in the {\it same} way $x_i \rightarrow \lambda x_i, t\rightarrow \lambda t$. Relatedly, the GCA does {\it not} admit a mass term as a central extension. Thus, in some sense, this symmetry describes "massless" or "gapless" non-relativistic theories, like the parent relativistic group but unlike the Schrodinger group.

 One of the most interesting feature of the GCA is its natural extension to an infinite dimensional symmetry algebra, somewhat analogous to the way the finite 2d conformal algebra of $SL(2,C)$ extends to two copies of the Virasoro algebra\footnote{Closely related infinite dimensional algebras have been studied in the context of statistical mechanical systems in \cite{Henkel06}. It would be interesting to study the precise connection as well as the potential realisations in statistical mechanics further.}. It is natural to expect this to be dynamically realized (perhaps partially) in actual systems possesing the finite dimensional Galilean conformal symmetry. This partial realization is actually observed in the non-relativistic Navier-Stokes equations\cite{Bagchi:2009my}. 

It has been known (see \cite{Duval:1993pe} and references therein) that there is a notion of a "Galilean isometry"  which encompasses the so-called Coriolis group of arbitrary time dependent (but spatially homogeneous) rotations and translations.  In this language, our infinite dimensional algebra is that of "Galilean conformal isometries".  It contains one copy of a Virasoro together with an $SO(d)$ current algebra (on adding the appropriate central extension). There has been interesting progress in this direction recently in \cite{Duval:2009vt}.

In a follow-up work \cite{Bagchi:2009}, we looked at representations and correlation functions of the GCA. Among other things, it was found that the form of the three point function was fixed upto a constant by the requirement of GCA invariance, like in the case of relativistic CFTs. However, in the case of the Schrodinger algebra, the three point function is arbitrary upto a function of a particular combination of variables. This is another indicator of the fact that the GCA is a more natural non-relativistic limit of the parent theory. For other related work on Galilean conformal algebras, see \cite{Alishahiha:2009np, Martelli:2009uc}.

A natural and immediate direction of interest is to try and generalize our construction to the supersymmetric case. The algebra was, as emphasised just above, obtained by taking a natural parametric limit of the relativistic conformal algebra. We thus expect that the GCA would be a sub-sector of all relativistic theories. Particularly, in the context of AdS/CFT, we would like to take the limit on the ${\mathcal{N}}=4$ Supersymmetric Yang-Mills theory. It is natural, in this context, to first try and extend the analysis of the algebra to include supersymmetry before we look to understand the details of the full field theory.  We would finally be interested in embedding the GCA in String Theory where we would need to realize supersymmetric configurations. 

The paper is organised as follows. In Sec 2, we revisit the bosonic GCA discussed earlier in \cite{Bagchi:2009my, Bagchi:2009} where we review the group contractions leading to the GCA and its infinite extension. In Sec 3, we construct, in detail, the ${\mathcal{N}}=1$ supersymmetric extension of the GCA, which we will refer to as SGCA. We also lift the SGCA to an infinite dimensional algebra and comment on the generalisation to higher ${\mathcal{N}}$. We conclude in Sec 4, with comments and a list of things which are to be done.

\bigskip

{\bf{Note added:}} While this paper was being readied for submission, two papers appeared \cite{Sakaguchi:2009de,deAzcarraga:2009ch} which have considered the finite supersymmetric extensions of the GCA.

\section{Bosonic Galilean Conformal Algebra}

\subsection{Contraction of the Relativistic Conformal Group}

We know that the Galilean algebra $G(d,1)$ arises as a contraction of the Poincare algebra
$ISO(d,1)$. Physically this comes from taking the non-relativistic scaling 
\be{nrelscal}
t \rightarrow t\,, \quad   x_i \rightarrow \epsilon x_i\,,
\ee
with $\epsilon \rightarrow 0$. This is equivalent to taking the velocities $v_i \sim \epsilon$ to zero
(in units where $c=1$).

Starting with the expressions for the Poincare generators ($\mu,\nu=0,1\ldots d$)
\be{mm}
J_{\mu\nu} = -(x_{\mu} \p_{\nu} - x_{\nu} \p_{\mu})\,, \qquad P_{\mu}=\p_{\mu}\,,
\ee
the above scaling gives us the Galilean vector field generators
\ben{galvec}
J_{ij}&=& -(x_i\p_j-x_j\p_i)\,, \qquad P_0=H= -\p_t\,, \cr
P_i &=& \p_i \qquad J_{0i}=B_i= t\p_i .
\een
They obey the following commutation relations (Galilean sub-algebra):
\ben{galalg}
[ J_{ij}, J_{rs} ] &=& so(d)\,, \cr
[J_{ij} , B_r ] &=& -(B_i {\delta}_{jr} - B_j {\delta}_{ir}) \,,\cr
[J_{ij},\, P_r] &=& -(P_i {\delta}_{jr} - P_j {\delta}_{ir})\,, \quad [J_{ij},\, H] = 0\,, \cr
 [B_i,B_j] &=& 0\,, \quad [P_i, P_j] =0\, ,\quad [B_i, P_j] =0\,, \cr
[H, P_i] &=& 0\,, \quad [H, B_i] = - P_i\,. 
\een

To obtain the Galilean Conformal Algebra, we simply extend the scaling \eq{nrelscal} to the rest of the generators of the conformal group $SO(d+1,2)$. Namely to 
\be{dkrel}
D = -(x\cdot\p) \qquad K_{\mu} = -(2x_{\mu}(x\cdot\p) -(x\cdot x)\p_{\mu})\,,
\ee
where $D$ is the relativistic dilatation generator and $K_{\mu}$ are those of special conformal transformations.
The non-relativistic scaling in \eq{nrelscal} now gives (see also \cite{Lukierski:2005xy})
\ben{nrelconf}
D &=& -( x_i \p_i + t \p_t) \,,\cr
K&=& K_0= -(2t x_i \p_i +t^2 \p_t)\,, \cr
K_i &=& t^2\p_i\,. 
\een

Note that the dilatation generator $D=-( x_i \p_i + t \p_t)$ 
is the {\it same} as in the relativistic theory. It scales space and time in the same way $x_i \rightarrow \lambda x_i , t \rightarrow \lambda t $. It is different from the dilatation generator $\tilde{D} =-(2t\p_t+x_i\p_i)$ of the Schrodinger group which scales space and time differently. 
Similarly, the temporal special conformal generator $K$ in \eq{nrelconf} is 
different from $\tilde{K} = -(tx_i\p_i+t^2\p_t)$.
Finally, we now have spatial special conformal transformations $K_i$ (which were not present in the Schrodinger algebra). Thus the generators of the Galilean Conformal Algebra are 
$(J_{ij}, P_i, H, B_i, D, K, K_i)$.

The other non-trivial commutators of the GCA are \cite{Lukierski:2005xy}
\ben{galconalg}
[K, K_i] &=&0\,, \quad [K, B_i]=K_i\,, \quad [K, P_i]= 2B_i\,, \cr
[J_{ij}, K_r] &=& -(K_i {\delta}_{jr} - K_j {\delta}_{ir})\,, \quad [J_{ij}, K] =0\,, \quad [J_{ij}\,, D]=0\,, \cr
[K_i,K_j] &=& 0\,, \quad [K_i, B_j]=0\,, \quad [K_i,P_j]=0\,, \quad [H, K_i] = -2B_i\,, \cr
[D, K_i] &=& -K_i\,, \quad [D , B_i]=0\,, \quad [D, P_i] = P_i\,,\cr
[D,H] &=& H\,,  \quad [H, K]= -2D\,, \quad [D, K]=-K\,.
\een

This can be contrasted with commutators of the corresponding relativistic generators:
\ben{poinconalg}
[K, K_i] &=&0\,, \quad [ K,  B_i]= K_i\,, \quad [ K,  P_i]= 2 B_i\,, \cr
[ J_{ij},  K_r] &=& -( K_i {\delta}_{jr} - K_j {\delta}_{ir})\,, 
\quad [ J_{ij},  K] =0\,, \quad [J_{ij},  D]=0\,, \cr
[K_i, K_j] &=& 0\,, \quad [ K_i,  B_j]= \delta_{ij} K\,, \quad [ K_i, P_j]
=2J_{ij} +2\delta_{ij}D \,,\cr 
[ H,  K_i] &=& -2 B_i\,, \quad [ D,  K_i] = - K_i\,, 
\quad [ D ,  B_i]=0\,, \quad [D,  P_i] =  P_i\,,\cr
[D, H] &=& H\,,  \quad [ H, K]= -2D, \quad [ D, K]=- K\,.
\een

The Schrodinger algebra and the GCA only share a common Galilean subgroup and are otherwise different. 
In fact, one can verify using the Jacobi identities for $(D,B_i, P_j)$ that the Galilean 
central extension in $[B_i,P_j]$ is {\it not} admissible in the GCA. This is another difference from
the Schrodinger algebra, which does allow for the central extension. 
Thus in some sense, the GCA is the symmetry  of a "massless" (or gapless) nonrelativistic system.

\subsection{The Infinite Dimensional Extended GCA}

The most interesting feature of the GCA is that it admits a very natural extension to an infinite 
dimensional algebra of the Virasoro-Kac-Moody type \cite{Bagchi:2009my}. To see this we denote 
\ben{rename}
L^{(-1)} &=& H, \qquad L^{(0)}=D, \qquad L^{(+1)}= K, \cr
M_i^{(-1)} &=& P_i, \qquad M_i^{(0)}=B_i, \qquad M_i^{(+1)}=K_i.
\een
The finite dimensional GCA which we had in the previous section can now be recast as
\ben{gcafinit}
[J_{ij}, L^{(n)}] &=&0 ,  \qquad [L^{(m)}, M_i^{(n)}] =(m-n)M_i^{(m+n)}, \cr
[J_{ij} , M_k^{(m)} ] &=& -(M_i^{(m)} {\delta}_{jk} - M_j^{(m)} {\delta}_{ik}), 
\qquad [M_i^{(m)}, M_j^{(n)}] =0, \cr
[L^{(m)}, L^{(n)}] &=& (m-n)L^{(m+n)}.
\een
The indices $m,n=0,\pm 1$
We have made manifest the $SL(2,R)$ subalgebra with the generators $L^{(0)}, L^{(\pm 1)}$. 
In fact, we can define the vector fields 
\ben{gcavec}
L^{(n)} &=& -(n+1)t^nx_i\p_i -t^{n+1}\p_t\,, \cr
M_i^{(n)} &=& t^{n+1}\p_i \,,
\een 
with $n=0,\pm 1$. These (together with $J_{ij}$) are then exactly the vector fields
in \eq{galvec} and \eq{nrelconf} which generate the GCA (without central extension). 

If we now consider the vector fields of \eq{gcavec} for {\it arbitrary} integer $n$, and also define
\be{Jn}
J_a^{(n)} \equiv J_{ij}^{(n)}= -t^n(x_i\p_j-x_j\p_i)\,,
\ee
then we find that this collection obeys the current algebra 
\ben{vkmalg}
[L^{(m)}, L^{(n)}] &=& (m-n)L^{(m+n)} \qquad [L^{(m)}, J_{a}^{(n)}] = -n J_{a}^{(m+n)}\,, \cr
[J_a^{(n)}, J_b^{(m)}]&=& f_{abc}J_c^{(n+m)} \qquad  [L^{(m)}, M_i^{(n)}] =(m-n)M_i^{(m+n)}. 
\een
The index $a$ labels the generators of the spatial rotation group $SO(d)$ and $f_{abc}$ are the
corresponding structure constants. 
We see that the vector fields generate a $SO(d)$ Kac-Moody algebra without any central terms. In addition to the Virasoro and current generators we also have the commuting generators 
$M_i^{(n)}$ which function like generators of a global symmetry. We can, for instance, consistently set these generators to zero. The presence of these generators therefore do not spoil the ability of the Virasoro-Kac-Moody generators to admit the usual central terms in their commutators. 

\section{Supersymmetric Extension of the GCA}

The main objective of this paper is to systemetically construct a supersymmetric extension of the GCA. To that end, we would first look to perform a contraction on the simplest ${\mathcal{N}}=1$ case. As in the bosonic case, we would look to implement the contraction at the level of the co-ordintes. We write down the superspace representations of the relativistic algebra and perform the contraction on the ordinary as well as the grassmann co-ordinates. 

\subsection{Non-relativistic Contraction in Superspace} 
We have seen that for the bosonic case, the non relativistic limit arises from \refb{nrelscal}. Now, we also need to take into account the grassmann co-ordinates $\theta$. We know that square of $\theta$ acts like an ordinary co-ordinate. Hence we expect the scaling of $\theta$ to go like $\sqrt{\e}$. The various linear combinations of the components of $\theta$ that scale differently should scale like $\sqrt{\e}$ and $1\over{\sqrt{\e}}$ respectively. Naive choices of the components would lead to incorrect answers like the vanishing of  $\{Q, \overline{Q} \}$. So, one needs to be careful while choosing the appropriate linear combinations of the components of the grassmann variable which would scale in the way mentioned above.

Along with \refb{nrelscal}, we choose to scale 
\be{thscale}
\t_+ \to {1\over{\sqrt{\e}}} \t_+\,, \quad \t_- \to \sqrt{\e} \t_-\,,
\ee 
where $\t_{\pm}$ are projections defined below:
\be{proj}
\t_{\pm} = {1\over 2} (1 \pm \gamma^0) \t = {1\over 2} \pmatrix{{{\mathbf{1}}_{2 \times 2} \,\  \pm \sigma^0} \cr {{ \pm \sigma^0 \,\ {\mathbf{1}}_{2 \times 2}}}} \pmatrix{\t_{\a} \cr \tb^{\dot \a}}\,.
\ee 

We list the conventions used for the spinor algebra in Appendix A. To make the process of contraction explicit, let us define variables with the different scaling behaviours as follows:
\ben{psichi}
\s_1 &=& {1 \over 2} (\t_1 - \tb_{\dot 2}), \quad \s_2 = {1 \over 2} (\t_2 + \tb_{\dot 1}) \quad \Rightarrow \s_{1,2} \to  {1\over{\sqrt{\e}}} \s_{1,2} \cr 
\c_1 &=& {1 \over 2} (\t_1 + \tb_{\dot 2}), \quad \c_2 = {1 \over 2} (\t_2 - \tb_{\dot 1}) \quad \Rightarrow \c_{1,2} \to  {\sqrt{\e}} \c_{1,2} 
\een

Let us briefly comment on the choice of this particular scaling. From the bosonic case, we know that the space part should scale in a way which is different from the time part in the non-relativistic limit. We would have a rotational symmetry present in the spatial part of the non-relativistic quantities of interest. Incorporating this feature in the supersymmetric case, it is necessary to look at spinors of $SO(3)$. The spinors we have described above are indeed spinors of $SO(3)$. Some more details can be found in Appendix A. 
There is another hint at what we want if we look at the non-relativistic limit of the Dirac equation. This is precisely the projection that gets rid of the negative energy states and projects onto the Pauli equation in the limit where we take the speed of light to infinity. 

\subsection{Fermionic Generators}

We would implement the above described scaling on the fermionic generators\footnote{Here we would be keeping track of factors of $\pm i$ which we had not taken into account in the purely bosonic subalgebra.}. We would look at particular combinations of the generators and perform the contraction in a way similar to the bosonic case. 

We look at the supersymmetry generators first. 
They are 
\be{susygen}
Q_{\a} = i {\p \over{\p \t^{\a}}} + \sigma_{\a \dot \a}^{\mu} \tb^{\dot \a} \p_{\mu}\,, \quad \overline{Q}_{\dot\a} = -i {\p \over{\p \tb^{\dot \a}}} + \t^{\a}\sigma_{\a \dot \a}^{\mu}\p_{\mu}\,.
\ee
where $\a , {\dot \a} = 1,2$.

More explicitly:
\ben{susygenexp}
Q_1 = i {\p \over{\p \t_2}} - \tb_{\dot 2} \p_t + \tb_{\dot 2} \p_3 - \tb_{\dot 1} \p_1 + i \tb_{\dot 1} \p_2\,, \quad  
Q_2 = -i {\p \over{\p \t_1}} + \tb_{\dot 1} \p_t + \tb_{\dot 1} \p_3 + \tb_{\dot 2} \p_1 + i \tb_{\dot 2} \p_2\,, \nonumber \\
\overline{Q}_{\dot 2} = i {\p \over{\p \tb_{\dot 1}}} - \t_{1} \p_t - \t_{1} \p_3 - \t_{2} \p_1 + i \t_{2} \p_2\,, \quad 
\overline{Q}_{\dot 1} = -i {\p \over{\p \tb_{\dot 2}}} + \t_{2} \p_t - \t_{2} \p_3 - \t_{1} \p_1 + i \t_{1} \p_2 \,.\nonumber
\een

In order to perform the contraction, we would need to take linear combinations of the above equations as shown below : 
\ben{recomb}
{\tilde {\Q}}^{+}_1 = Q_1 - \overline{Q}_{\dot 2}\, , \quad
{\tilde {\Q}}^{+}_2 = Q_2 + \overline{Q}_{\dot 1}\,, \nonumber \\
{\tilde {\Q}}^{-}_1 = Q_1 + \overline{Q}_{\dot 2}\, , \quad
{\tilde {\Q}}^{-}_2 = Q_2 - \overline{Q}_{\dot 1} \,.
\een
Expressing them in the variables $\s_a$ and $\c_a$ defined before, we find
\ben{qq}
{\tilde {\Q}}^{+}_a &=&  i \e^{ab}\p_{\c_b} + 2 \psi_{a} \p_t + 2 \c_{b} \sigma^j_{ab} \p_j \,,   \cr
 {\tilde {\Q}}^{-}_a &=& i \e^{ab}\p_{\psi_b} - 2 \c_{a} \p_t-2 \psi_b  \sigma^j_{ab} \p_j \,,\nonumber
\een
where $a  = 1,2$.

We now perform the contraction by scaling $t$, $x^i$, $\s_{a}$ and $\c_{a}$ and choosing redefined generators in the way below :
\ben{contr}
{\Q}^{+}_a &=& \lim_{\e \to 0} \e^{1/2} {\tilde {\mathcal{Q}}}^{+}_a = i \e^{ab}\p_{\c_b} + 2 \psi_{a} \p_t + 2 \c_{b} \sigma^j_{ab} \p_j \,,   \cr 
{\Q}^{-}_a &=& \lim_{\e \to 0} \e^{3/2} {\tilde {\mathcal{Q}}}^{-}_a = -2 \psi_b  \sigma^j_{ab} \p_j\,.
\een

The anticommutators of the algebra involving ${\mathcal{Q}}^{-}_{1,2}$ are all zero. The non-zero anticommutators are given by 
\be{susycom} 
\{ \Q^{+}_1, \Q^{+}_2 \} =- 4 i\p_3, \quad \{ \Q^{+}_1, \Q^{+}_1 \} = 4 i (\p_1 - i \p_2), \quad \{ \Q^{+}_2, \Q^{+}_2 \} = - 4 i (\p_1 + i \p_2) .
\ee 

Now we turn our attention to the other fermionic generators of the relativistic super-conformal group,- the super-conformal transformations $S$. 

The S-supersymmetry generators are
\ben{ssusygen}
S_{\a} = -i \e^{\dot{\beta}\dot{\gamma}} (\sigma_{\mu})_{\a \dot{\gamma}} x^{\mu}_{(+)} \t^{\beta} \sigma^{\nu}_{\beta \dot{\beta}} \partial_{\nu} + 2i(\t\t)\p_{\a} + \e^{\dot{\beta}\dot{\gamma}} (\sigma_{\mu})_{\a \dot{\gamma}} x^{\mu}_{(-)} \overline{\p}_{\dot \beta}\,, \nonumber \\
\overline{S}_{\dot\a} = -i \e^{\beta \gamma} (\sigma_{\mu})_{\gamma \dot{\a}} x^{\mu}_{(-)} \overline{\t}^{\dot \beta} \sigma^{\nu}_{\beta \dot{\beta}} \partial_{\nu} - 2i(\overline{\t} \overline{\t}) \overline{\p}_{\dot \a} + \e^{\beta \gamma} (\sigma_{\mu})_{\gamma \dot \a} x^{\mu}_{(+)} \p_{\beta} \,,\nonumber
\een
where we have defined $x^{\mu}_{(\pm)} = x^{\mu} \pm i \t \sigma^{\mu} \tb$.

As in the case of the $Q$-generators, we take linear combinations of the above equations as follows:
\ben{recomb1}
{\tilde {S}}^{+}_1 = S_1 - \overline{S}_{\dot 2} \,, \quad 
{\tilde {S}}^{+}_2 = S_2 + \overline{S}_{\dot 1}\,, \nonumber \\
{\tilde {S}}^{-}_1 = S_1 + \overline{S}_{\dot 2}\, ,  \quad 
{\tilde {S}}^{-}_2 = S_2 - \overline{S}_{\dot 1} \,.
\een 

We now perform the contraction by scaling $t$, $x^i$, $\s_{a}$ and $\c_{a}$ and choosing redefined generators in the way below :
\ben{contr1}
{\mathcal{S}}^{+}_a = \lim_{\e \to 0} \e^{1/2} {\tilde {S}}^{+}_a &=& (2i t \c_{b} -8\c_{1} \c_{2}\psi_{b}) \sigma^{j}_{ab} \p_{j} + 2i \sigma^{j}_{ab} \sigma^{k}_{bc} \psi_{c} x^j \p_k
+ \psi_{a}  (2it - 8 \e^{bc}\psi_{b} \c_c )\p_t \cr
&& - 2i \psi_{a} \psi_{b} \p_{\psi_{b}} + 4i \psi_{a} \c_b \p_{\c_{b}} - 6i \e^{ab}\e^{cd} \psi_{c} \c_{d}\p_{\c_{b}} - t \e^{ab} \p_{\c_{b}} \,,\crcr
{\mathcal{S}}^{-}_a = \lim_{\e \to 0} \e^{3/2} {\tilde {S}}^{-}_a &=& 2i t \psi_{b} \sigma^{j}_{ab}  \p_{j} +8 \psi_{1}\psi_{2}\c_{b}  \sigma^{j}_{ab} \p_{j} - 6i\psi_{1}\psi_{2} \e^{ab} \p_{\c_{b}}\,,
\een
where $a  = 1,2$.

Again, the only non-zero anticommutators involve the ${\mathcal{S}}^{+}_a$ generators as shown below:
\be{s-alg}
\{{\mathcal{S}}^{+}_1, {\mathcal{S}}^{+}_2 \} = 4 {\mathcal{K}}_3, \quad \{{\mathcal{S}}^{+}_1, {\mathcal{S}}^{+}_1 \} = -4 ({\mathcal{K}}_1 -i {\mathcal{K}}_2), 
 \quad \{{\mathcal{S}}^{+}_2, {\mathcal{S}}^{+}_2 \} = 4 ({\mathcal{K}}_1 + i {\mathcal{K}}_2).
\ee 
We should mention that the bosonic generators now also have fermionic pieces. The details can be found in Appendix B. The algebra of the bosonic generators, as expected, remains the same with these additional pieces. Along with all the usual bosonic generators, there are also the extra R-symmetry generators which rotate the fermionic generators. For the ${\mathcal{N}}=1$ case at hand, this is just a single generator, representing the $U(1)$ R-symmetry. (Again, you can find more details in Appendix B.) 

\subsection{Algebra}
We list the algebra here but omitting the purely bosonic subalgebra. One should also note that the commutator of $\mathcal{A}$ with any bosonic generator is zero.

The non-zero anticommutators of the fermionic generators are given by
\ben{sgcaferm}
\{ \Q^+_1, \Q^+_1 \} &=& -4(P_1 -iP_2), \,\ \{ \Q^+_2, \Q^+_2 \} = 4(P_1 +iP_2), \,\ \{ \Q^+_1, \Q^+_2 \} = 4 P_3, \cr 
\{ \S^+_1, \S^+_1 \} &=& -4({\mathcal{K}}_1 -i{\mathcal{K}}_2), \,\ \{ \S^+_2, \S^+_2 \} = 4({\mathcal{K}}_1 +i{\mathcal{K}}_2), \,\ \{ \S^+_1, \S^+_2 \} = 4 {\mathcal{K}}_3, \cr 
\{ \S^+_1, \Q^+_1 \} &=& 4i (B_1 - iB_2), \,\ \{ \S^+_1, \Q^+_2 \} = -4 i B_3 - 12 {\mathcal{A}}, \cr 
\{ \S^+_2, \Q^+_1 \} &=& -4 i B_3 + 12 {\mathcal{A}}, \,\ \{ \S^+_2, \Q^+_2 \}= -4i (B_1 + iB_2) .
\een

The commutators of $\Q_a^{\pm}, \S_a^{\pm}$ with $P_i, {\mathcal{K}}_i, B_i$ are as follows:
\ben{qs}
[P_i, \Q_a^{\pm}] &=& 0, \,\ [P_i, \S_a^{+}] = -\sigma^i_{ab} \Q_b^{-},\,\ [P_i, \S_a^{-}] = 0, \cr
[{\mathcal{K}}_i, \Q_a^{+}] &=& -\sigma^i_{ab} \S_b^{-} ,\,\ [{\mathcal{K}}_i, \Q_a^{-}]= 0, \,\ [{\mathcal{K}}_i, \S_a^{\pm}] = 0, \cr
[B_i, \Q_a^{+}] &=& {i\over 2} \sigma^i_{ab} \Q_b^{-} , \,\ [B_i, \Q_a^{-}] = 0, \cr
[B_i, \S_a^{+}] &=& {i\over 2} \sigma^i_{ab} \S_b^{-} , \,\ [B_i, \S_a^{-}] = 0.
\een 

The commutators of $\Q_a^{\pm}, \S_a^{\pm}$ with the angular momentum generators $J_{i}$ are given by
\be{qsrot}
[J_i, \Q_a^{\pm}] = -{1\over 2}\sigma^i_{ab} \Q_b^{\pm}, \quad [J_i, \S_a^{\pm}] = -{1\over 2}\sigma^i_{ab} \S_b^{\pm},
\ee
which tell us that the fermionic generators transform as spinors of $SO(3)$, the three-dimensional rotation group.

Finally, the commutators of $\Q_a^{\pm}, \S_a^{\pm}$ with $H, {\mathcal{K}}, D, {\mathcal{A}}$ are as follows:
\ben{qsrest}
[H,  \Q^{\pm}_a] &=& 0 , \,\ [H,  \S^{+}_a] = \Q^{+}_a , \,\ [H,  \S^{-}_a] = -\Q^{-}_a,\cr
[{\mathcal{K}} , \Q^+_a] &=& -\S^+_a , \,\ [{\mathcal{K}} , \Q^-_a] = \S^-_a , \,\  [{\mathcal{K}} , \S^{\pm}_a]=0 , \cr
[D, \Q_a^{\pm}] &=& -{i\over 2} \Q_a^{\pm}, \,\ [D, \S_a^{\pm}] = {i\over 2} \S_a^{\pm}, \cr
[{\mathcal{A}}, {\Q}^{+}_a] &=& - {1\over 2} \Q^{-}_a, \,\ [{\mathcal{A}}, {\Q}^{-}_a] =0,  \cr
[{\mathcal{A}}, {\S}^{+}_a] &=& {1\over 2} \S^{-}_a, \,\ [{\mathcal{A}}, {\S}^{-}_a] =0.
\een

\subsection{Infinitely extended SGCA}

We have seen in Sec(2.2) that the bosonic GCA admitted an infinite dimensional extension. This was one of the very interesting aspects of the GCA. As we have systematically employed an analogous non-relativistic limit on the supersymmetric version of the bosonic conformal algebra, we would expect that similar infinite dimensional extensions are valid even in this case. To that end, we would now re-write the finite dimensional contracted algebra in a suggestive form. \\
We define:
\be{infg}
G_{-1/2} = \pmatrix{ G^{+a}_{-1/2} \cr G^{-a}_{-1/2}}= \pmatrix{i\Q^+_{a} \cr -i\Q^-_{a}}, \quad 
G_{1/2} = \pmatrix{ G^{+a}_{1/2} \cr G^{-a}_{1/2}} = \pmatrix{ \S^+_{a} \cr \S^-_{a}}.
\ee 
Remembering the definitions of $L_n, M^i_n, J_{ij}$ from Sec(2.2), we can re-write the finite dimensional superconformal algebra in the following way\footnote{Here we return to a convention devoid of $i$ for the purely bosonic subalgebra to compare with the results in the infinite bosonic algebra. More explicitly, the $H, B_i, K_i, J_{ij}$ used here can be obtained from those defined in Appendix B and used in Sec(3.3), by multiplying with $(-i)$, whereas the $D, K, P_i$ used here can be obtained by multiplying with $i$.}: 
\ben{SGCA}
[L_m, L_n] &=& (m-n)L_{m+n} , \,\ [M_n^i, M_m^j] = 0 , \,\ [L_m, M_n^i] = (m-n)M^i_{m+n},  \cr
[L_n, G_r^{\pm a}] &=& ({n\over 2} - r) G_{n+r}^{\pm a} , \,\  [M^i_n, G_r^{+a}] =  (r-{n\over 2}) \sigma^i_{ab} G_{n+r}^{-b}, \,\ [M^i_n, G_r^{-a}] = 0,  \cr
\{ G^{+a}_r, G^{+b}_s \} &=& 4i(\sigma^i \e)_{ab} M_{r+s}^i -12i f_1(r,s) \e^{ab} {\mathcal{A}}_{r+s}, \,\ \{ G^{-a}_r, G^{\pm b}_s \} = 0,
\een 
for $n=0, \pm 1$ and $r=\pm {1\over2}$.

At the level of the algebra, we can continue \refb{SGCA} for all integral values of $n,m$ and all half integral values of $r,s$. The commutators of the supercharges of the finite algebra with the $L_n, M_n^i$ for arbitrary $n$ generate the higher supercharges. We see that to fit with the Jacobi identity we need to promote ${\mathcal{A}}$ to have a Virasoro index\footnote{We would like to thank Ashoke Sen for this valuable input.}. Let us try and derive a consistent infinite lift for ${\mathcal{A}}$ so that the above algebra closes. The first thing to note is that the symmetry of the total fermionic anticommutator and the antisymmetry of $\e^{ab}$ forces the function $f_1(r,s)$ to be antisymmetric in $r,s$. Let us take the simplest function and try and build a consistent infinite dimensional extension of the contracted algebra, which is
\be{f1}
f_1(r,s) = r-s \, .
\ee 
To generate the rest of the ${\mathcal{A}}_n$ algebra, let us look at the Jacobi identity involving $G^{+a}_r, G^{+b}_s, {\mathcal{A}}_n$ : 
\ben{GGA}
&& [ \{G_r^{+a}, G_s^{+b} \} , {\A}_n ] - \{ [G_s^{+b}, \A_n], G_r^{+a} \} + \{ [\A_n, G_r^{+a}], G_s^{+b} \} = 0\,, \\ 
&\Rightarrow& 4i (\sigma^i \e)_{ab} [M^i_{r+s}, \A] -12i \e^{ab} (r-s) [ \A_{r+s}, \A_n ]= 0\,,  \nonumber 
\een 
where we have used $[\A_n, G^+_r] = f_2 (n,r) G^-_{n+r}$ . 

We find that
\be{MA-AA}
[M^i_m, \A_n] = 0, \quad [\A_m, \A_n] =0 
\ee 
is a consistent choice. The first commutator can be further motivated by the fact that the vector index on the RHS must come from $M^i_{n^{\prime}}$ and not $\sigma^i_{ab} \A_{n^{\prime}}$ as $\A_{n^{\prime}}$ does not have any $a$ or $b$ index to contract with the indices coming from $\sigma^i_{ab}$. Then the RHS would at most have factors of $M^i_{m+n}$. Any natural definition of  $[\A_m, \A_n]$ would not generate $M^i_{n^{\prime}}$ on the RHS. And hence, to satisfy \refb{GGA}, both must be given by \refb{MA-AA}. 

We can now look at the Jacobi identity of $L_m, L_n, \A_{m^{\prime}}$ to arrive at a choice for $[L_n,\A_m]$ which fits with the finite algebra. The simplest choice is
\be{LA}
[L_n, \A_m] = - m \A_{m+n} \, .
\ee 
The rest of the algebra of $\A_n$, and indeed the full contracted finite case, also follows from various Jacobi identities. We list a simple consistent choice for the infinite lift of the rest of the algebra:
\be{rest-alg}
 [\A_n, G_r^{+a}] = {1\over 2} G_{n+r}^{-a}, \quad[\A_n, G_r^{-a}] =0,\quad [J^i, G_r^{\pm a}] = {i \over 2} \sigma^i _{ab} G_{r}^{\pm b}, \quad [J^i, \A_n]= 0 .
\ee 
Together with \refb{SGCA}, \refb{MA-AA}, and \refb{LA}, \refb{rest-alg} constitutes the infinite dimensional extension of the non-relativistic superconformal algebra\footnote{Here we cannot give the rotation generators $J^i$'s an infinite lift as discussed in Appendix C. The OPE analysis of Appendix C also suggests that this simple choice is the unique infinite dimensional extension consistent with the finite part for this particular contraction.}. 

The interesting thing to note here is that the supercharge anticommutators generate the $M_n^i$ and not the $L_n$ as would be the usual expectation from the bosonic algebra. The reason for this counter-intuitive behaviour can be traced back to the fact that we chose to scale the fermionic generators in a way which meant that the $SL(2,R)$ part always dropped out of the fermionic anti-commutators. The above algebra looks structurally like the usual superconformal algebra in 2 dimensions. This is an infinitely extended ${\mathcal{N}}=1$ Super Galilean conformal algebra.

There might be a possible way to extract the usual supeconformal algebra in 2 dimensions by choosing to contract the fermionic generators in a different way. One has to make sure that the $SL(2,R)$ part is the one that remains after scaling and not the part with the vector indices. This would be a different non-relativistic limit of the parent super-conformal algebra and hence an inequivalent Super-Galilean conformal algebra. It might be interesting to explore this in more detail. The classification of the possible supersymmetric GCAs is also something worth pondering about.

\subsection{Generalization to higher ${\mathcal{N}}$} 

The generalization to extended SUSY is immediate. For ${\mathcal{N}}>1$ superconformal algebras, the difference with the ${\mathcal{N}}=1$ case is that the number of fermionic generators increases. (With each of the $Q$ and $S$, a label '$p$' will now be attached.) This, in turn, implies that the R-symmetry will be enhanced for these algebras. We can use constructions very similar to what we have described above to arrive at the non-relativistic extended superconformal algebras. The linear combinations used earlier would just need to be augmented by the extra internal index '$p$' and the R-symmetry generator $\mathcal{A}$ gets promoted to ${\mathcal{A}}_{pq}$ (for example, $SO(6)$ symmetry for ${\mathcal{N}}=4$).

There is a point to note here. Unlike in the case of extended supersymmetry algebras, extended superconformal algebras don't allow for central extensions in the fermionic generators. This is clear if one looks at the Jacobi identities. For example, if we want to put a central term in the $\{ {\overline{Q}},{\overline{Q}} \}$ anticommutator, then we can check using the Jacobi identity of $\{P, S, {\overline{Q}} \} $ that this cannot be consistent. So, in the extended superconformal algebras, we would not have to worry about additional central terms coming from the relativistic algebra when we are looking to perform the non-relativistic contraction. The process of contraction and the contracted algebra are just the same as in the ${\mathcal{N}}=1$ with extra indices fitted wherever required. 

The infinite dimensional lift can also be implemented along the lines mentioned in the previous subsection. Except for extra indices, the rest of the construction remains the same.

\section{Concluding Remarks}

In this paper, we have systematically derived a non-relativistic limit for superconformal algebras. We used a simple representation of the ${\mathcal{N}}=1$ superconformal algebra and took a limit on that by demanding that the linear combinations of the spinors we look at would be spinors of $SO(3)$. We also found a way to embed this finite dimensional algebra in an infinitely extended algebra, along the lines similar to the bosonic construction. The suprise there was that instead of the appearance of the expected infinite ${\mathcal{N}}=1$ superconformal algebra in $d=2$, we found a close cousin of that algebra. The extension to extended superconformal algebras was immediate. These are supersymmetric versions of the bosonic Galilean Conformal Algebra studied in \cite{Bagchi:2009my,Bagchi:2009}. It is interesting that one has a simple framework in which all types of superconformal algebras can be handled in the non-relativistic limit. We also mentioned the fact that there might be other contractions of the superconformal algebra, which realise different types of super-extended GCAs. In this context, we would like to mention that in \cite{Gomis}, the authors looked at similar contractions of the relativistic super-conformal group $PSU(2,2|4)$ while looking at non-relativistic limits of AdS/CFT correspondence from the world-sheet point of view. The linear combination of spinors used there is motivated by kappa-symmetry of the relativistic string action and seem to be different to our limit\footnote{The recent work \cite{Sakaguchi:2009de} mentioned in the introduction, seems more in this direction.}. 

There exists a large literature on the supersymmetric extensions of the Schrodinger algebra; some of the recent works include \cite{Duval:1993hs}--\cite{Lee:2009mm}. Among these, \cite{Sakaguchi:2008rx,Sakaguchi:2008ku} rely on clever re-writings of the relativistic algebras to get at the super-schrodinger algebra which exists as a subgroup inside the relativistic one. It is interesting to contrast their approach to ours and it might be worth trying to understand if we could adopt such an approach to arrive at the SGCA.

There are numerous avenues left to explore in this context. Construction of representations and correlation functions for the supersymmetric case, along the lines of \cite{Bagchi:2009} is an immediate step. Now that we understand the non-relativistic scaling in a super-conformal setting, we are better equipped to deal with the main subject of interest, viz. ${\mathcal{N}}=4$ SYM. The primary objective of the supersymmetric extension of the GCA that we have looked at in this paper is to build a platform from which we can understand the symmetries of the ${\mathcal{N}}=4$ SYM. Once we understand how to take this systematic non-relativistic limit of ${\mathcal{N}}=4$ SYM, we might be able to isolate some interesting and tractable sector of the theory, somewhat analogous to the plane wave sector in the BMN limit. 

The bulk dual of the GCA proposed in \cite{Bagchi:2009my} was a novel Newton-Cartan like limit of $AdS_{d+2}$ with a base $AdS_2$ and $R^d$ fibres. The GCA emerges as the asymptotic isometry\cite{Brown:1986nw} of this Newton-Cartan structure. A better understanding of the boundary theory is also a step in understanding the bulk. Taking the non-relativistic limit on the ${\mathcal{N}}=4$ SYM would be a first step in trying to understand the novel bulk-boundary dictionary in this case. Now that we understand how to extend the GCA to SGCA, using the contraction on the ${\mathcal{N}}=4$ SYM should become clearer. We are at present looking at this and other related questions. 

\subsection*{Acknowledgements}
The authors would like to thank Rajesh Gopakumar and Ashoke Sen for many helpful discussions and for carefully reading and commenting on the manuscript. We would also like to record our indebtedness to the people of India for their generous support for fundamental enquiries.


\section*{Appendices}
\appendix

\section{Spinors: Conventions and Choice}

In this appendix, we provide some details of first the conventions and then why we choose to scale the spinors in the way we have done in the paper. 

\subsection{Conventions}

For the calculations in this paper, we have followed the notation of Wess-Bagger.
\ben{conventions}
\sigma^{\mu} &=& (\sigma^0, \sigma^i), \quad {\overline{\sigma}}^{\mu} = (\sigma^0, - \sigma^i) \nonumber\\
\sigma^0 &=& \pmatrix{-1 & 0 \cr 0 & -1}, \quad  \sigma^1 = \pmatrix{0 & 1 \cr 1 & 0}, \quad \sigma^2 = \pmatrix{0 & -i \cr i & 0}, \quad \sigma^3 = \pmatrix{1 & 0 \cr 0 & -1} \nonumber\\
(\sigma^{\mu \nu})_{\a}^{\,\ \beta} &=& -{1\over 4} (\sigma^{\mu} {\overline{\sigma}}^{\nu} - \sigma^{\nu} {\overline{\sigma}}^{\mu})_{\a}^{\,\ \beta}, \quad 
({\tilde{\sigma}}^{\mu \nu})^{\dot \a}_{\,\ \dot \beta} = -{1\over 4} ({\overline{\sigma}}^{\mu} {{\sigma}}^{\nu} - {\overline{\sigma}}^{\nu} {\sigma}^{\mu})^{\dot \a}_{\,\ \dot \beta} \nonumber \\
\e^{12} &=& -\e^{21}= \e^{\dot 1 \dot 2}= - \e^{\dot 2 \dot 1} = 1 \nonumber \\
\e_{12} &=& -\e_{21}= \e_{\dot 1 \dot 2}= - \e_{\dot 2 \dot 1}=-1 \nonumber
\een  

We have used the $a,b$ indices running over $1,2$, when we have dealt with the components of the non-relativistic $SO(3)$ spinors, and it does not matter whether they have been written as subscripts or superscripts. More explicitly, 
\be{}
\e^{ab} =\e_{ab}\,, \quad (\sigma^i)_{ab} = (\sigma^i)^{ab} =(\sigma^i)^a_{\,\,\,b}=(\sigma^i)_a^{\,\,\,b}\,. \nonumber
\ee

\subsection{Choice of spinors}

In the non-relativistic limit there is a sharp distinction between time and space which was not present in the relativistic theory. So, as we have already stressed, when we look at the symmetries of quantities of interest in a non-relativistic setting, we would expect a rotational invariance in the spatial part only. In $d=3+1$, when we consider fermions, we would thus choose to work with spinors of $SO(3)$.

From the commutators of the $\Q$'s with the $J_{ij}$ generators, it is clear that under spatial rotations, they have the transformation properties of the spinors of $SO(3)$.

More explicitly, let us look at how the combinations mentioned in the paper form the two components of an $SU(2)$ or an $SO(3)$ spinor. \\
Under $SO(3)$ transformations: 
\be{so3}
\delta \t_{\a} = (\sigma_{ij})_{\a}^{\,\ \beta} \t_{\beta} w^{ij}, \quad  \delta \tb_{\dot \beta} = - \tb_{\dot \a}({\tilde{\sigma}}_{ij})^{\dot \a}_{\,\ \dot \beta} w^{ij}.
\ee 
Now, 
\be{v}
\sigma_{ij} =  {i \over 2 } \e_{ijk} \sigma^k \Rightarrow w^{ij} \sigma_{ij} = \pmatrix{c & a-ib \cr a+ib & -c} .
\ee
So, we get
\ben{ttb}
\delta \t_1 &=&  c\t_1 + (a-ib) \t_2, \quad  \delta \t_2 = (a+ib) \t_1 - c \t_2 \cr 
\delta \tb_{\dot 1} &=&  -(a+ib)\tb_{\dot 2} - c \tb_{\dot 1}, \quad \delta \tb_{\dot 2} =  c\tb_{\dot 2} - (a-ib) \tb_{\dot 1}.
\een
Hence, 
\ben{diff}
\delta (\t_1- \tb_{\dot 2}) = c (\t_1-\tb_{\dot 2}) + (a-ib) (\t_2+\tb_{\dot 1}) \,,\cr 
\delta (\t_2+ \tb_{\dot 1}) = (a+ib) (\t_1-\tb_{\dot 2}) - c (\t_2+\tb_{\dot 1})\,.
\een 
Or more compactly,
\be{t+}
\t_{+} = \pmatrix{\t_1- \tb_{\dot 2} \cr \t_2+ \tb_{\dot 1} } \Rightarrow \delta{\t_+} = ( \hat{n}.\overrightarrow{\sigma}) \t_+\,.
\ee

This shows that the linear combinations that we work with are indeed $SO(3)$ spinors. 

\section{Bosonic Generators}
As already mentioned in Sec(3.2), the bosonic generators in the supersymmetric case will get contributions from the fermionic coordinates. In this appendix, we list the various bosonic generators that have extra pieces and note how the generators need to be rescaled in the non-relativistic limit. As expected, the scaling behaviour is the same as in the case of the purely bosonic algebra. 

We first derive explicit expressions for the angular momentum and boost generators, which are obtained from the $J_{ij}$ and the $J_{0i}$ components respectively, of the Lorentz generators $J_{\mu \nu}$ given by
\be{lorentz}
 J_{\mu \nu} = i (x_{\nu} \p_{\mu}-x_{\mu} \p_{\nu} + \sigma_{\mu \nu}^{\a \beta} \t_{\a} \p_\beta-{\tilde{\sigma}}_{\mu \nu}^{\dot \a \dot \beta} \overline{\t}_{\dot \a}\overline{\t}_{\dot \beta})\,.
\ee 
We now perform contraction by scaling $x^{i}$, $t$, $\psi_{a}$ and $\chi_{a}$.
The $J_{ij}$ generators are invariant under this scaling and are explicitly given by
\ben{ang-mom}
J_{ij} &=& i (x_j \p_i - x_i \p_j) + {1 \over 2}\e_{ijk} \left \lbrace \psi_{a}(\sigma^k)^{T}_{ab} \p_{\psi_{b}}  + \c_a (\sigma^k)^{T}_{ab}\p_{\c_b} \right \rbrace \,.
\een
For notational convenience, we define
\be{J}
J_i = {1 \over 2}\e_{ijk} J_{jk}\,.
\ee
For the $J_{0i}$ generators, however, we have to choose redefined operators as shown below:
\ben{boost}
B_{i} &=& \lim_{\e \to 0} \e J_{0i} = it \p_i + {i \over 2} \psi_{a} (\sigma^i)^{T}_{ab}  \p_{\c_b}\,.   
\een
\\
The dilatation generator is given by the expression
\ben{dil}
D &=& i x^{\mu} \p_{\mu} + {i \over 2} (\t^{\a} \p_{\a} + \overline{\t}^{\dot \a} \overline{\p}_{\dot \a}) \cr 
&=& it \p_{t} + i x^{j} \p_{j}+ {i \over 2} (\psi_{a} \p_{\psi_{a}} + \c_a \p_{\c_a}) \,.
\een
We can see clearly that $D$ does not scale when we perform contraction by scaling $x^{i}$, $t$, $\psi_{a}$ and $\chi_{a}$. Hence we need not redefine it.

The remaining bosonic generators are given by the temporal and spatial special conformal transformation generators, obtained from the $K_0$ and the $K_i$ components, respectively, of the relativistic special conformal transformation generator $K_{\mu}$ given by
\ben{Kmu}
K_{\mu} &=& -4i \left \lbrace x_{\mu} x^{\nu} + (\t \sigma_{\mu} \overline{\t}) (\t \sigma^{\nu} \overline{\t})\right \rbrace \p_\nu + 2i \left \lbrace x^{\nu} x_{\nu} + 2(\t \t)(\overline{\t}\overline{\t})\right \rbrace \p_\mu \cr
&& + 4i \e^{\beta \gamma} (\sigma^{\nu} \overline{\sigma}_{\mu})_\gamma \, ^\a \t_\a x_\nu \p_\beta \,.
\een

We perform contraction by scaling $x^{i}$, $t$, $\psi_{a}$ and $\chi_{a}$.
For these generators we have to choose redefined operators as shown below :
\ben{Ki}
{\mathcal{K}}_{i} &=& \lim_{\e \to 0} \e K_{i} = i(t^2 + 16 \psi_1 \psi_2  \c_1 \c_2)\p_i + it \psi_a (\sigma^i)^T_{ab} \p_{\c_b} +4 \psi_1 \psi_2 \c_a (\sigma^i)^T_{ab} \p_{\c_b}\,, \cr
{\mathcal{K}} &=& \lim_{\e \to 0}  K_{0} = -i (t^2 - 16 \psi_1 \psi_2  \c_1 \c_2)\p_t -2it x^j \p_j -it (\psi_a \p_{\psi_a}  + \c_a \p_{\c_a}) \cr
&& +4\psi_1 \psi_2  \c_a \p_{\psi_a} -4 \c_1 \c_2 \psi_a \p_{\c_a} -i x^j \psi_a (\sigma^j)^T_{ab} \p_{\c_b}\,.
\een

The R-symmetry generator is given by the expression
\ben{r1}
A &=& {1 \over 2} (\t^{\a} \p_{\a} -\overline{\t}^{\dot \a} \overline{\p}_{\dot \a}) \cr \cr
&=& {1 \over 2} (\psi_{a} \p_{\c_{a}}  + \c_a \p_{\psi_a}) \,.
\een
Performing the contraction by scaling $x^{i}$, $t$, $\psi_{a}$ and $\chi_{a}$, we find that we need to redefine it as
\ben{r}
\mathcal{A} &=& \lim_{\e \to 0} \e A = {1 \over 2} \psi_{a} \p_{\c_{a}}  \,.
\een

Just for the sake of completeness, we give here the expressions used in Sec(3.3) for the Hamiltonian $H$ and the momentum generators $P_i$, which do not have any piece contributed by the fermionic coordinates. They are:
\be{mom}
H= -i\p_t \,, \quad P_i = -i \p_i\,.
\ee

\section{OPE Analysis}
We consider the holomorphic currents $T(z), M^i(z), A(z), G^{\pm a}(z)$, where $z$ denotes the two-dimensional complex plane. Now we regard the generators $L_n, M^i_n, A_n, G^{\pm a}_r$ of our infinite-dimensional algebra as the Laurent coefficients or mode operators of the Laurent expansion of the above holomorphic currents respectively. Then the (anti)commutators of the mode operators can be found by the standard Operator Product Expansion (OPE) and contour integral method.

If we could give $J^i$ also an infinite lift, it would have to be promoted to a holomorphic current like the others. However, for the choice $[J^i_n, A_m]=0$, the Jacobi identity for $J^i_n,G^{+ a}_r,G^{+ b}_s$ is not satisfied for $n \neq 0$, which tells us that $J^i$ cannot be given an infinite lift in this framework.

However, let us adopt an OPE analysis to see if we could have modified our infinite algebra, assuming $J^i$ can be promoted to a holomorphic current $J^i (z)$ in this modified framework. We will then find all the (anti)commutators and fix the infinite algebra by requiring that all the Jacobi identities should be satisfied, and the finite part should coincide with \refb{SGCA} and have the commutators of ${\mathcal{A}}_0$ with $L_0, L_{\pm 1},M_0^i,M_{\pm 1}^i, J^i_0 $ equal to zero.

The OPEs are given by
\ben{ope}
T(z_1)T(z_2) &\sim & {2 T(z_2) \over (z_1-z_2)^2}  + {\p_{z_2}T(z_2) \over (z_1-z_2)}  \,,\cr\cr
T(z_1)M^i (z_2) &\sim & {a_1 J^i(z_2) \over (z_1-z_2)^3} + {a_2 \p_{z_2}J^i(z_2) \over (z_1-z_2)^2} + {2 M^i(z_2) \over (z_1-z_2)^2} + {\p_{z_2}M^i(z_2) \over (z_1-z_2)} \,,\cr\cr
T(z_1)J^i (z_2) &\sim & {J^i(z_2) \over (z_1-z_2)^2}  + {\p_{z_2}J^i(z_2) \over (z_1-z_2)} + {a_3 M^i(z_2) \over (z_1-z_2)} \,,\cr\cr
T(z_1) \A(z_2) &\sim & {a_4 \over (z_1-z_2)^3}+{\A(z_2) \over (z_1-z_2)^2}  + {\p_{z_2}\A (z_2) \over (z_1-z_2)} + {a_5 T(z_2) \over (z_1-z_2)} \,,\cr\cr
T(z_1)G^{\pm a} (z_2) &\sim & {{3 \over 2} G^{\pm a}(z_2) \over (z_1-z_2)^2}  + {\p_{z_2}G^{\pm a}(z_2)\over (z_1-z_2)}  \,,\cr\cr
M^i(z_1)G^{+ a} (z_2) &\sim & -{{3 \over 2}\sigma^i_{ab}G^{+ b}(z_2) \over (z_1-z_2)^2}  - {\p_{z_2}\sigma^i_{ab}G^{+ b}(z_2) \over (z_1-z_2)}  \,,\cr\cr
M^i(z_1)G^{- a} (z_2) &,& M^i(z_1)M^i(z_2), \, \,\, {\mathcal{A}}(z_1){\mathcal{A}}(z_2) \,\,\,\sim  \mbox{non-singular} \,,\cr\cr
G^{+ a}(z_1)G^{+ b}(z_2)&\sim & -{2 \times 12i \e^{ab} {\mathcal{A}}(z_2) \over (z_1-z_2)^2}  + {4i (\sigma^i \e)_{ab}M^i(z_2)-12i \e^{ab} \p_{z_2}{\mathcal{A}}(z_2) \over (z_1-z_2)} \,,\cr\cr
 M^i(z_1){\mathcal{A}}(z_2) &\sim &  \mbox{non-singular}\,,\,\,J^i(z_1){\mathcal{A}}(z_2) \sim {a_6 \over (z_1-z_2)} J^i(z_2)\,,\cr\cr
{\mathcal{A}}(z_1) G^{+ a} (z_2) &\sim & {G^{- a} (z_2)\over 2(z_1-z_2)} + a_7{G^{+ a} (z_2)\over (z_1-z_2)} \,,\,\,{\mathcal{A}}(z_1) G^{- a} (z_2) \sim a_8{G^{- a} (z_2)\over (z_1-z_2)} + a_9{G^{+ a} (z_2)\over (z_1-z_2)}\,, \nonumber
\een
where $a_l=0$ ($1\leq l \leq 9$) for our algebra of Sec(3.4).

Now let us see if we can allow for non-zero values of the $a_l$'s. One can check that we have allowed for all possible terms in the OPEs when one considers the weights of the various holomorphic fields and the index structure involving $i$ and $a$.

With the above consideration, we have the following modified commutators:
\ben{}
[L_m, M_n^i]&=& (m-n)M_{m+n}^i + {a_1 \over 2}m(m+1)J_{m+n}^i - a_2 (m+1)(m+n+1)J_{m+n}^i \,,\cr
[L_m, J_n^i]&=& -n J_{m+n}^i + a_3 M_{m+n}^i \,,\cr
[L_m, {\mathcal{A}}_n^i]&=& -n {\mathcal{A}}_{m+n}^i + {a_4 \over 2} m(m+1) \delta_{m+n,0}+a_5 L_{m+n}\,,\cr
[J_n^i,{\mathcal{A}}_m]&=& a_6 J_{m+n}^i \,,\,\,\,[{\mathcal{A}}_n,G^{+ a}_r ]= {1 \over 2}G^{- a}_{n+r} + a_7G^{+ a}_{n+r}\,,\,\,\,[{\mathcal{A}}_n,G^{- a}_r ]= a_8 G^{- a}_{n+r} + a_9 G^{+ a}_{n+r}\, . \nonumber
\een
But now one can easily check that consistency with the finite part of the algebra sets $a_1, a_2, a_3, a_6,a_7,a_8,a_9$ to zero, while the Jacobi identity for $\{ L_m,L_n, \A_p \}$ sets $a_4$ and $ a_5$ to zero.
The above analysis suggests that we cannot give the $J^i$'s an infinite lift.


\end{document}